\documentclass[12pt]{article}
\usepackage{amssymb}
\usepackage{epsf}
\usepackage{epsfig}
\usepackage{psfrag}
\usepackage{rotating}
\renewcommand{\baselinestretch}{1.2}

\newcommand{\query}[1]{\marginpar{%
  \vskip-\baselineskip 
  \raggedright\footnotesize
  \itshape\hrule\smallskip#1\par\smallskip\hrule}}
\newcommand{\removequeries}{\renewcommand{\query}[1]{}}

\newcommand{\lsim}{
\mathrel{\hbox{\rlap{\hbox{\lower4pt\hbox{$\sim$}}}\hbox{$<$}}}}
\newcommand{\gsim}{
\mathrel{\hbox{\rlap{\hbox{\lower4pt\hbox{$\sim$}}}\hbox{$>$}}}}

\newcommand{\vus}{|V_{us}|}

\newcommand{\be}{\begin{equation}}
\newcommand{\ee}{\end{equation}}
\newcommand{\bi}{\begin{itemize}}
\newcommand{\ei}{\end{itemize}}

\newcommand{\IM}{{\rm Im}}
\def\kpn{K^+\rightarrow\pi^+\nu\bar\nu}
\def\klpn{K_{\rm L}\rightarrow\pi^0\nu\bar\nu}

\newcommand{\Br}{{\cal B}}
\newcommand{\Kpnn}{{K\to\pi\nu\bar\nu}}

\begin{document}
\removequeries

\begin{flushright}
TUM-HEP-583/05\\
BNL-73846-2005-JA \\
\end{flushright}

\vspace*{0.3truecm}

\begin{center}
\boldmath
{\Large{\bf $\klpn$ as a Probe of New Physics}} 
\unboldmath
\end{center}

\vspace{0.0 truecm}

\begin{center}
{\bf Douglas Bryman,${}^a$  Andrzej J. Buras,${}^b$ \\
Gino Isidori${}^c$, and Laurence Littenberg${}^d$} \\
\vspace{0.4truecm}
${}^a$ {\sl Department of Physics and Astronomy, 
University of British Columbia, \\
6224 Agricultural Road, Vancouver, BC V6T1Z1 Canada}  \\
\vspace{0.2truecm}
${}^b$ {\sl Physik Department, Technische Universit\"at M\"unchen,
D-85748 Garching, Germany} \\
\vspace{0.2truecm}
${}^c$ {\sl INFN, Laboratori Nazionali di Frascati, Via E. Fermi 40, I-00044 Frascati, Italy} \\
\vspace{0.2truecm}
${}^d$ {\sl Brookhaven National Laboratory, PO Box 5000, Upton, NY 11973, USA} \\
\end{center}

\vspace{0.6cm}

\begin{abstract}
\vspace{0.2cm}\noindent
We summarize the theoretical virtues of the rare $\Kpnn$  decays 
and emphasize the unique role of
$\klpn$ in probing the nature of physics beyond the Standard Model, 
in particular concerning possible new sources of 
CP violation and flavor-symmetry breaking. A brief summary of the 
prospects for the measurement of the $\klpn$ rate 
is also given.
\end{abstract}

%
%
%

\vspace{1.0 true cm}

\section{Introduction}\label{sec:intro}
\setcounter{equation}{0}
The rare decays of $K$ and $B$ mesons play an important role in the search 
for the underlying mechanism  of 
flavor dynamics and in particular in the search for the 
origin of CP violation~\cite{Schladming}. 
Among the many $K$ and $B$ decays, the rare 
decays $\kpn$ and 
$\klpn$ are very special as 
their branching ratios can be computed to an exceptionally high degree of 
precision, not matched by any other flavor-changing
neutral-current (FCNC) process involving quarks.
While the theoretical uncertainties in the branching ratios of
prominent FCNC processes, such as $B\to X_s\gamma$  and
$B\to X_s\mu^+\mu^-$, amount 
to $\pm 10\%$ or larger, 
the  irreducible theoretical uncertainty in $\Br(\klpn)$ amounts 
to only 1-2$\%$~\cite{BB2,MU98,BB98,GBGI}. 
The non-negligible charm contribution 
leads to a slightly larger theoretical error in
the case of $\Br(\kpn)$: $\pm 8\%$ at the NLO level~\cite{BB98,BB3}, which will 
soon be reduced significantly thanks to both the NNLO calculation
of the leading partonic amplitude~\cite{BUGOHANI} and the 
recent progress in the evaluation of long-distance effects~\cite{IMS}.
A recent very detailed review of $\kpn$ and 
$\klpn$ in the Standard Model and in its most popular extensions has 
been presented in~\cite{BSU}, where the usefulness of both processes 
for the determination of the SM parameters and in the search for the physics 
beyond the SM has been emphasized and summarized. 
Other theoretical reviews can be found in~\cite{Gino03}, while the prospects 
for the measurements of these decays have been summarized 
in~\cite{Bryman,Littenberg}.

According to the detailed analysis in \cite{BSU}, the present
predictions for the branching ratios of the two decay modes 
within the SM are 
\begin{equation}
\label{SMkp+}
\Br(\kpn)_{\rm SM}=
(7.8 \pm 1.2)\cdot 10^{-11}~, 
\ee
\be\label{SMkl0}
 \Br(\klpn)_{\rm SM}=
(3.0 \pm 0.6)\cdot 10^{-11}~,
\end{equation}  
where a good fraction of the error
($\pm15\%$ and $\pm 20\%$, respectively)
is due to parametric uncertainties 
(CKM angles and quark masses). 
Thanks to the foreseen theoretical progress in the evaluation 
of $K\to \pi\nu\bar\nu$ amplitudes and, especially, the expected 
improvement in the determination of the CKM parameters
from BaBar, Belle, CDF, D0, and other experiments,
these predictions should reach  the $\pm 5\%$ level, 
or better, in a few years.
This accuracy cannot be matched by any other loop-induced 
process in the field of meson decays.

On the experimental side, the AGS E787 and E949 collaborations at Brookhaven 
observed the decay $\kpn$ \cite{Adler970,Adler02,E949} finding three events
so far.  The resulting branching ratio is
\be\label{EXP1}
\Br(\kpn)=
(14.7^{+13.0}_{-8.9})\cdot 10^{-11}~.
\ee
The central value of this measurement is substantially higher
than the SM prediction in (\ref{SMkp+}). However, taking 
into account the substantial uncertainties in~(\ref{EXP1}), 
as well as theoretical and parametric errors, the present result
is consistent with the SM expectation.

So far, the best direct experimental information on the 
$\klpn$ mode is the  KTeV bound: $\Br(\klpn)<5.9 \cdot 10^{-7}$
\cite{E799}, which is about four orders of magnitude 
above the SM expectation.
A more stringent constraint can be derived using the 
information on the charged mode and isospin symmetry \cite{GRNR}:
\begin{equation}\label{GNBound}
\Br(\klpn) \lsim \frac{ \tau_{K_L} }{\tau_{K^+}  } \Br(\kpn)
\end{equation}
which through (\ref{EXP1})  gives
\begin{equation}\label{GNBounda}
\Br(\klpn) < 1.4 \cdot 10^{-9} \qquad (90\% {\rm C.L.}).
\end{equation}
As discussed in \cite{GRNR}, this bound is valid in
virtually any extension of the SM.
By comparing this model-independent 
bound and the SM prediction in (\ref{SMkp+}),
it is clear that there is still much room for new physics 
in $\klpn$. As we shall discuss in the following,
this corresponds to unexplored regions in the  
parameter space of several realistic 
new physics scenarios. But even if the experimental 
measurement of $\Br(\klpn)$ were found in agreement 
 with the SM expectation with a small relative error, 
this information would translate into a {\em unique} and precious insight 
about the CP and flavor structure
of any extension of the SM. These features makes 
the experimental search for $\klpn$, at the SM level and below, 
a win--win opportunity.

\boldmath
\section{$\kpn$ and $\klpn$ within the SM}
\begin{figure}[t]
\vspace{0.10in}
\begin{center}
\epsfysize=1.1in
\epsffile{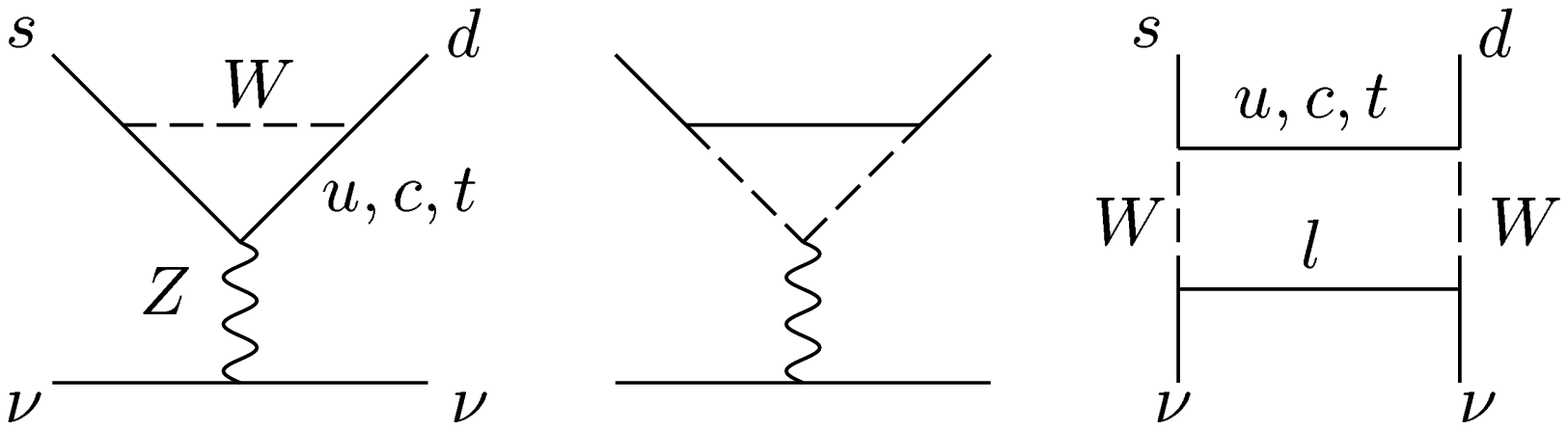}
\hspace{0.5 cm}
\epsfysize=1.2in
\epsffile{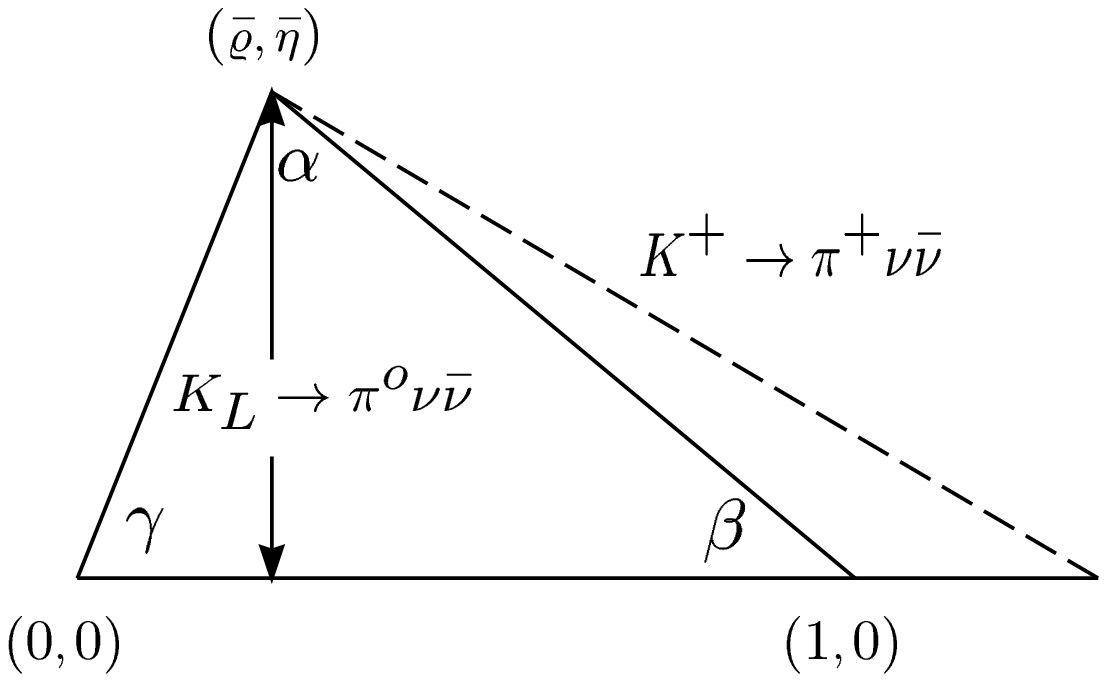} 
\end{center}
\vspace{-0.5 cm}
\hspace{5 cm} a)
\hspace{7 cm} b)
\caption{Leading Feynman diagrams relevant to $K\to\pi\nu\bar\nu$ decays (a);
CKM unitarity triangle from $K\to\pi\nu\bar\nu$ (b). \label{KPKLUT}}
\end{figure}

\setcounter{equation}{0}
\unboldmath
The main reason for the exceptional theoretical cleanness of 
$\kpn$ and $\klpn$~\cite{littenberg:89} decays is the fact that
--within the SM-- these 
processes are mediated by electroweak amplitudes 
of $O(G_F^2)$, which exhibit a power-like GIM mechanism~\cite{GIM}
(see Fig.~\ref{KPKLUT}). 
This property implies a severe suppression of 
non-perturbative effects~\cite{GBGI,IMS,RS,LW,FalkLP}.\footnote{~Higher-order 
electroweak effects on the leading $O(G_F^2)$ amplitude have 
also been computed and found to be safely negligible~\cite{BB5}.}
By comparison, it should be noted that typical 
loop-induced amplitudes relevant to meson decays 
are of $O(G_F\alpha_s)$ (gluon penguins) or 
$O(G_F\alpha_{\rm em})$ (photon penguins), 
and have only a logarithmic-type GIM mechanism,
which implies a much less severe suppression of 
non-perturbative effects. A related important virtue,
following from this peculiar electroweak structure,
is the fact that $K\to\pi\nu\bar\nu$
amplitudes can be described in terms of a single effective operator:
\be
Q_{sd}^{\nu\nu} = \bar{s} \gamma^\mu  (1-\gamma_5) d ~ \bar{\nu} \gamma_\mu   (1-\gamma_5) \nu~.
\label{eq:QnL}
\ee
The hadronic matrix elements of  $Q_{sd}^{\nu\nu}$
relevant to  $K\to\pi\nu\bar\nu$ amplitudes 
can be extracted directly from the well-measured 
$K^+\to\pi^ 0e^+\nu$ decays, including isospin 
breaking corrections~\cite{MP}.

In view of these features, the measurements of the two 
$K\to\pi\nu\bar\nu$ branching ratios can be translated 
--within the SM-- into  precise information on the CKM 
matrix and, in particular, on the so-called  
CKM unitarity triangle  \cite{BB96}. 
As shown in Fig.~\ref{KPKLUT}, 
$\Br(\klpn)$ determines the height of this triangle, 
while $\Br(\kpn)$ determines one of its sides.
Assuming that both branching ratios will be  known to within $\pm 10\%$,
one  expects the following accuracy on various quantities of interest~\cite{BSU}:
\be
\sigma(\sin 2\beta)=\pm 0.05, \quad
\sigma({\rm Im}\lambda_t)=\pm 5\%, \quad
\sigma(|V_{td}|)= \pm 7\%, \quad \sigma(\gamma)=\pm 11^\circ~. 
\ee
where $\lambda_t=V^*_{ts}V_{td}$, with $V_{ij}$ being the elements 
of the CKM matrix and $(\beta,\gamma)$ the angles of the unitarity 
triangle (see Fig.~\ref{KPKLUT}).
With the measurements of the branching ratios at the $\pm 5\%$ level 
these estimates change to 
\be
\sigma(\sin 2\beta)=\pm 0.03, \quad
\sigma({\rm Im}\lambda_t)=\pm 3\%, \quad
\sigma(|V_{td}|)= \pm 4\%, \quad \sigma(\gamma)=\pm 6^\circ~.
\ee
Further details can be found in \cite{BSU}.

It is worth stressing that the determination of CKM
parameters via $K\to\pi\nu\bar\nu$ decays is mainly an
efficient way to compare the measured value of these 
clean FCNC transitions with other clean tree-level
mediated or loop-induced observables. 
Since the loop-induced observables are potentially affected by  
non-standard contributions, this comparison offers a powerful 
tool to constrain or identify new-physics effects.
For instance, one of the most interesting studies which could be 
performed with experimental data on the two branching ratios, 
is a test of the so-called ``golden relation" \cite{BB4}:
\be\label{G1}
(\sin 2\beta)_{\psi K_S} = (\sin 2\beta)_{\pi\nu\bar\nu}~.
\ee
Here the right-hand side stands for the value of 
$\sin 2\beta$ determined from the two $K\to\pi\nu\bar\nu$
rates (see Fig.~\ref{KPKLUT}), while the left-hand side
denotes the corresponding value extracted at $B$ factories 
from the time-dependent CP asymmetry in $B^0_d\to \psi K_S$. 
This relation is not only a very powerful tool to falsify the SM, 
but also a useful handle to discriminate among 
different new-physics scenarios.

A key feature of the $\klpn$ mode is the fact that 
it proceeds through a pure loop-induced {\em direct}-CP-violating
amplitude \cite{littenberg:89}. Within the SM, its rate 
gives the cleanest determination of $\IM\lambda_t$, 
or the combination of Yukawa couplings which control the amount 
of CP violation in the model \cite{Jarlskog}. We can indeed write~\cite{BB96}
\begin{equation}\label{imlta}
\IM\lambda_t=1.39\cdot 10^{-4} 
\left[\frac{\vus}{0.224}\right]
\left[\frac{1.53}{X(x_t)}\right]
\sqrt{\frac{\Br(\klpn)}{3\cdot 10^{-11}}}~,
\end{equation}
where $X(x_t)$, with $x_t=m_t^2/M^2_W$, is the 
leading coefficient function of the operator $Q_{sd}^{\nu\nu}$
(according to the present value of the top-quark mass, 
$X(x_t)=1.53\pm 0.04$). Contrary to the $\klpn$
case,  in essentially all other $K$ and $B$ meson decays the extraction 
of loop-induced direct-CP-violating amplitudes is subject 
to sizable (if not huge) non-perturbative effects. 
This is, for instance, the case of the currently popular
direct CP-violating studies in non-leptonic two--body $B$ decays,
both those involving time-dependent distributions and those 
involving branching ratios and charge asymmetries.
Either the processes are tree-level dominated
(and thus naturally insensitive to new-physics effects)  
or it is very difficult to determine their 
direct-CP-violating phases with good 
theoretical control.

\boldmath
\section{The unique role of $\klpn$ in probing physics beyond the SM}
\label{sect:BSM}
\setcounter{equation}{0}
\unboldmath
\subsection{Preliminaries}
There are several reasons why the decay 
$\klpn$ plays a special role in the investigation 
of possible physics beyond the SM. 
While some of these reasons have been already emphasized in the literature, 
we would like to stress here a few points 
that we find particularly important:
\begin{itemize}
\item
The clean theoretical character of $\klpn$ (similarly of $\kpn$)
remains valid in essentially all 
extensions of the SM, whereas this is generally not the case for non-leptonic 
two-body B decays used to determine the CKM parameters through CP 
asymmetries and/or other strategies. While several mixing induced CP 
asymmetries  in non-leptonic B decays within the SM are essentially free 
from hadronic uncertainties, as the latter cancel out due to the dominance 
of a single CKM amplitude, this is often not the case in extensions of the SM 
in which the amplitudes receive new contributions with different weak phases 
implying no cancellation of hadronic 
uncertainties in the relevant observables.  
\item
The theoretically clean determinations of CP-violating phases in non-leptonic 
$B$ decays are based on tree level decays that are quite generally 
insensitive to new physics in the decay amplitudes and can be affected 
only by new phases in $B^0-\bar B^0$ mixing. In $\klpn$ the contributions 
from the CP violation in $K^0-\bar K^0$ mixing are by several orders
of magnitude smaller than the direct CP violation in the decay amplitude
\cite{littenberg:89} and consequently the direct CP violation in the SM 
and in its extensions can be tested here in a very clean environment.
Due to the different structure of the corresponding electroweak 
amplitudes, new-physics effects could be quite different
in direct- and indirect-CP-violating amplitudes (see e.g.~\cite{GI_radcor}).  
The former are poorly tested so far, because of  the sizable non-perturbative 
uncertainties which affect non-leptonic process both in $B$ and $K$ 
decays. This implies that there is still much room in the new-physics 
parameter space which can only be explored by means of $\klpn$.
\item
One of the most popular (and well motivated) scenarios
about the flavor structure of physics beyond the SM 
is the so-called {\it Minimal Flavor Violation} (MFV) hypothesis~\cite{UUT,DGIS}.
Within this framework (which can be regarded as the 
most {\em pessimistic} case for new-physics effects in rare decays), 
flavor- and CP-violating interactions are induced only by 
terms proportional to the SM Yukawa couplings. 
This implies that deviations from the SM 
in FCNC amplitudes rarely exceed the O$(20\%)$ 
level, or the level of irreducible theoretical 
errors in most of the presently available observables,
although model independently effects of order $50\%$ 
cannot be excluded at present \cite{MFV05}.
Moreover, theoretically clean quantities such as  
$a_{\rm CP}(B\to J/\Psi K_S)$ and $\Delta M_{B_d}/\Delta M_{B_s}$, 
which measure only ratios of FCNC amplitudes,  
turn out to be insensitive to new-physics effects. 
Within this framework, the need for additional 
clean and precise information on  FCNC transitions
is therefore even more important. A precise 
measurement of $\Br(K_L\to \pi^0 \nu \bar \nu)$ 
would offer a unique opportunity in this respect.
\end{itemize}

\subsection{General parameterization and phenomenological considerations}
An important consequence of the first item in the above list,
is the fact that in most SM extensions the new physics 
contributions in $\kpn$ and $\klpn$ 
can be parameterized in a model-independent manner 
by just two parameters, the magnitude 
and the phase of the Wilson coefficient of the operator 
$Q_{sd}^{\nu\nu}$ 
in Eq.~(\ref{eq:QnL}).\footnote{~For a discussion about the scenarios 
where this parameterization does not hold, see \cite{GIM2}.} 
More explicitly, we can encode all the new-physics effects around and 
above the electroweak scale into an effective Hamiltonian of the type
($\lambda_t=V^{\ast}_{ts}V_{td}$)
\be
{\cal H}_{eff} (M_W^2) = 
\frac{G^2_F M_W^2 }{ 2 \pi^2 }~\lambda_t~ X~ Q_{sd}^{\nu\nu} 
 ~+~ [\mbox{\footnotesize non-FCNC\ terms}] ~+~ {\rm h.c.}
\label{eq:heff} 
\ee
where the short-distance function \cite{BRS}
\be
\label{NX}
X=|X|e^{i\theta_X}
\ee
is such that the SM case corresponds to $|X| \to X(x_t) =1.53\pm 0.04$ and $\theta_X \to 0$. 
The important virtue of the $K\to\pi\nu\bar\nu$ system is that
$|X|$ and $\theta_X$ can be extracted from $\Br(\klpn)$ and $\Br(\kpn)$ 
without hadronic uncertainties, while the function 
$X$ can be calculated in any extension of the SM
within a perturbative framework.

\begin{figure}[t]
\begin{center}
\includegraphics[width=10 cm]{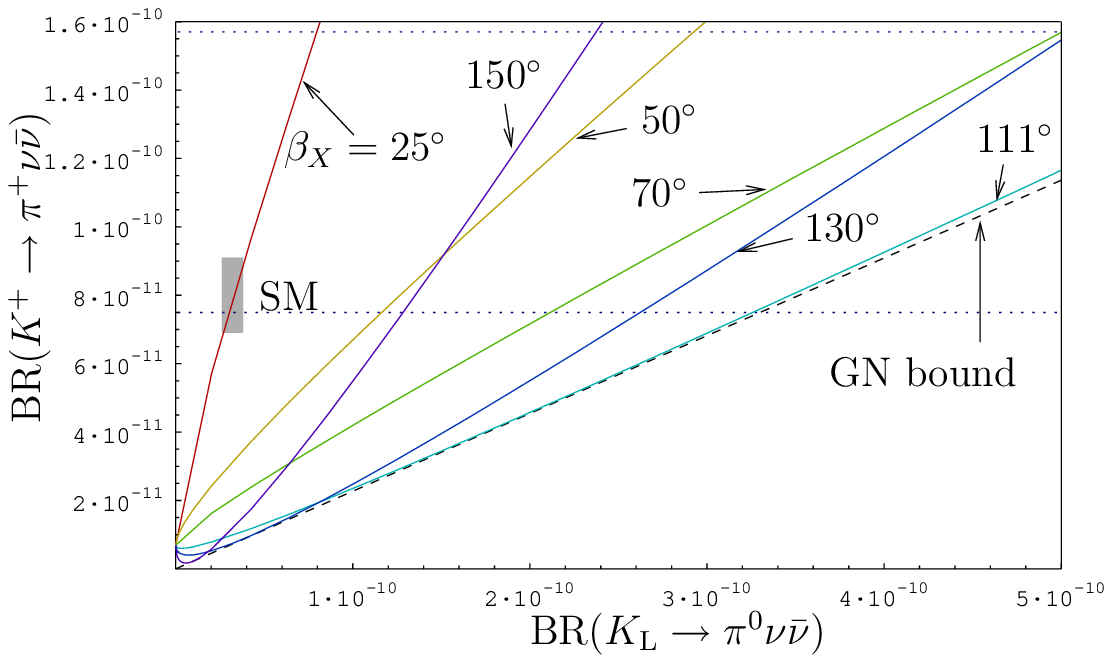}
\includegraphics[width=9.5 cm]{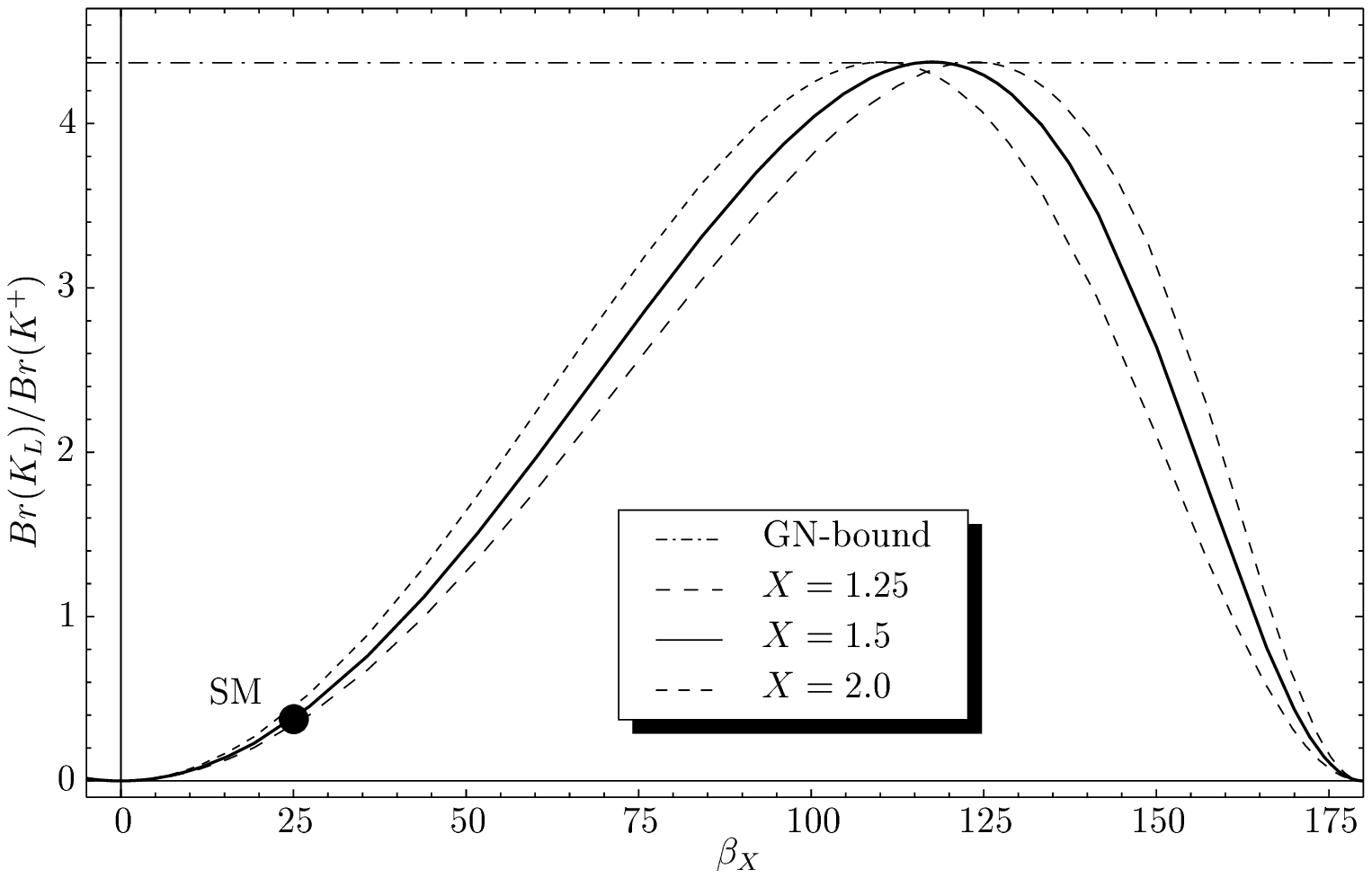}
\end{center}
\vspace{-0.5 truecm}
\caption{Up: $\Br(\kpn)$ as a function of $\Br(\klpn)$
for various values of $\beta_X=\beta-\theta_X$ \cite{BFRS-PRL}; 
the dotted horizontal lines indicate 
the lower part of the experimental range (\ref{EXP1}); the SM 
range and the bound (\ref{GNBound}) are also indicated. 
Down: ratio of charged and neutral branching ratios as a
function of $\beta_X$ for $|X|=1.25,~1.5,~2.0$ \cite{BSU}. \label{KLKPR}}
\end{figure}

The modulus of $X$ is directly constrained by $\Br(\kpn)$, 
which  is not very sensitive to $\theta_X$, while
$\Br(\klpn)$ strongly depends on $\theta_X$. A non-vanishing value 
of $\theta_X$ would signal the presence of extra CP-violating 
phases in $K\to\pi\nu\bar\nu$ amplitudes in addition to the 
standard CKM phase. In general, we can write 
\be
\frac{{\Br}(\klpn)}{{\Br}(\klpn)_{\rm SM}}=
\left|\frac{X}{X_{\rm SM}}\right|^2
\left[\frac{\sin(\beta-\theta_X)}{\sin(\beta)}\right]^2~,
\label{eq:KLratio}
\ee
where $\beta\approx 23^\circ$ is the standard angle of the unitarity triangle 
(or the phase of the CKM factor $V^{\ast}_{ts}V_{td}$). At present, the first factor 
on the right-hand side of Eq.~(\ref{eq:KLratio}) is constrained by 
the experimental 
data on $\Br(\kpn)$ to be smaller than $\approx 7$ (at 90\% C.L.).
However, even with an infinitely precise and completely 
SM result for $\Br(\kpn)$, one would still have 
much room for possible enhancements in $\Br(\klpn)$ 
due to the second factor, which could be as large as 
$\approx 6$. Combining these two possible enhancement factors, 
one recovers the present large potential for enhancement
of $\Br(\klpn)$ over its SM prediction, as also derived by 
the comparison of (\ref{SMkl0}) and (\ref{GNBounda}).
The pattern of the two $\Kpnn$ branching ratios as a function of $\theta_X$
is illustrated in Fig.~\ref{KLKPR}. Note, in particular, that the 
ratio of the two modes depends very mildly  
on $|X|$ and provides the ideal tool to extract 
the non-standard CP-violating phase $\theta_X$.

\medskip 

The $X$ function has been defined assuming 
the SM normalization (electroweak couplings + CKM factors) 
for the $Q_{sd}^{\nu\nu}$ operator. In principle, 
the non-standard effects could originate through a very 
different type of dynamics, such that this normalization 
would not be the most natural one. To estimate the 
new-physics sensitivity of $\Br(\klpn)$ on pure dimensional grounds, 
we can denote by $\lambda_{sd}/\Lambda_{\rm NP}^2$ the overall coefficient 
of the extra (non SM) contribution to the $Q_{sd}^{\nu\nu}$ operator.
If the generic dimensionless coupling $\lambda_{sd}$ is of $O(1)$,
it follows that a measurement of $\Br(\klpn)$
with $10\%$--$20\%$ accuracy allows probing   
new-physics scales well above 100 TeV. 
To be more precise, a measurement of $\Br(\klpn)$ with 
central value equal to the SM prediction and relative 
precision $p=\sigma\Br/\Br$, allows setting the following 
90\% CL bound on the scale of the operator: 
\be
\Lambda_{\rm NP}/\sqrt{{\rm Im} \lambda_{sd} }
 ~ > ~ \left[ \frac{G^2_F M_W^2 }{ 2 \pi^2 } \IM\lambda_t  X(x_t)  \right]^{-1/2} 
\left( 0.64 ~p \right)^{-1/2}
~\stackrel{p= 0.1}{\longrightarrow} ~ 1280~{\rm TeV}~ !
\label{eq:maxL}
\ee    
This remarkably high scale corresponds 
to the effective mass of new particles only in the 
extreme scenarios where new physics effects contribute to $\klpn$
at the tree level and all the relevant couplings are $O(1)$. 
As discussed in the following sections, the effective couplings 
are usually much smaller in more realistic models.
Nonetheless, even in the most pessimistic case, 
namely within MFV models, a $10\%$ measurement of 
$\Br(\klpn)$ allows probing new-physics scales 
well above the electroweak scale.

\subsubsection*{An explicit example}  
As pointed out in \cite{BFRS-PRL}, a scenario with a large 
phase $\theta_X\approx -90^\circ$ and a slightly enhanced $|X|$, 
has an interesting phenomenological motivation: assuming this
effect is flavor-universal, it would provide a much better fit 
of recent $B\to\pi K$ data from $B$ factories. 
According to this hypothesis, one would find 
\begin{equation}\label{NPkpnr}
\Br(\kpn)=
(7.5 \pm 2.1)\cdot 10^{-11}~,\quad \Br(\klpn)=
(3.1 \pm 1.0)\cdot 10^{-10}~,
\end{equation}
to be compared with the SM predictions in 
(\ref{SMkp+}) and (\ref{SMkl0}).

Apart from its phenomenological motivation, this explicit example
is useful for illustrating two important points:
\begin{itemize}
\item{} The values of $|X|$ and $\theta_X$ of this scenario
have been derived by fitting  a $10-20~\%$ deviation in the branching 
ratios of 
$B\to\pi K$ decays. This small effect in $B$ physics,
translates into an order of magnitude enhancement of 
$\Br(\klpn)$ over its SM estimate. This happens because $K\to\pi\nu\bar\nu$
amplitudes are completely dominated by short-distance electroweak 
dynamics and thus are very sensitive to possible non-standard 
effects above the electroweak scale. On the contrary, 
short-distance effects in non-leptonic $B$ decays are 
largely {\em diluted} by sizable long-distance 
contributions, which are insensitive to physics
above the electroweak scale. 
\item{} ${\Br}(\klpn)$ is naturally more sensitive to new physics 
than $\Br(\kpn)$. In particular, in this specific case
$\Br(\kpn)$ does not significantly
differ from the SM estimate. This happens because the enhancement of $|X|$ 
and the effects of large $\theta_X$, while being constructive 
in the $\Br(\klpn)$ case, compensate each other in 
$\Br(\kpn)$. 
\end{itemize}
It is worth stressing that, in spite of the 
phenomenological character of the analysis of \cite{BFRS-PRL}, 
such a configuration can be realized within consistent extensions of the SM.
In particular, as noted first in~\cite{CI}, and as confirmed 
by more recent detailed analyses~\cite{BCIRS,GMSSM}, 
a scenario of this type can be explicitly 
realized within low-energy supersymmetric
extensions of the SM.

\subsection{Testing specific models: the MFV hypothesis}
As we have seen in the previous section, the $\klpn$ decay 
is in principle sensitive to new physics up to very high scales.
However, this is true only in non-standard scenarios 
where the additional contributions to  $\Kpnn$ amplitudes
do not respect the strong CKM suppression present in the SM 
and are not governed by the GIM mechanism.
A similar behavior occurs in many other FCNC transitions, although 
the maximal sensitivity reachable in $B$ decays is substantially
smaller than the one in (\ref{eq:maxL}). For this reason,  
when discussing new-physics effects in rare FCNC processes,
it is very convenient to distinguish two basic scenarios: 
i) models with new sources of CP violation and flavor mixing; 
ii) models where, at the electroweak scale, these symmetries are 
effectively broken only by terms proportional to the (SM) 
Yukawa couplings. The latter is usually called hypothesis   
of Minimal Flavor Violation (MFV) \cite{UUT,DGIS}. 
As shown in \cite{DGIS}, this hypothesis can be formulated in 
a consistent way (in terms of an effective field theory), even without 
specifying the details of the new-physics model. It can also be shown 
that this hypothesis is the most pessimistic framework for rare decays:
given the Yukawa interaction breaks CP invariance and induces flavor mixing 
already within the SM, we cannot impose a more restrictive 
symmetry-breaking pattern beyond the SM~\cite{DGIS}.

The consequences of the MFV hypothesis for $\Kpnn$
decays have been discussed by several authors
(see \cite{BSU,DI} and references therein),
both in general and in specific frameworks where 
this hypothesis can naturally be implemented
(such as low-energy supersymmetry, universal extra-dimensions, 
little-Higgs models, etc.). On general grounds,
the MFV hypothesis forces $\Kpnn$ amplitudes to be 
proportional to the CKM factor $\lambda_t$. 
Thus, in these models the new-physics scale probed by  $\klpn$
is in the few TeV range, as can easily be understood by 
setting ${\rm Im} \lambda_{sd}={\rm Im} \lambda_t$
in (\ref{eq:maxL}).
Within all SM extensions which provide a natural solution 
to the hierarchy problem, this is the natural scale 
for new physics to show up.

\begin{figure}[t]
\vspace{-0.3 true cm}
\begin{center}
\includegraphics[width=12cm]{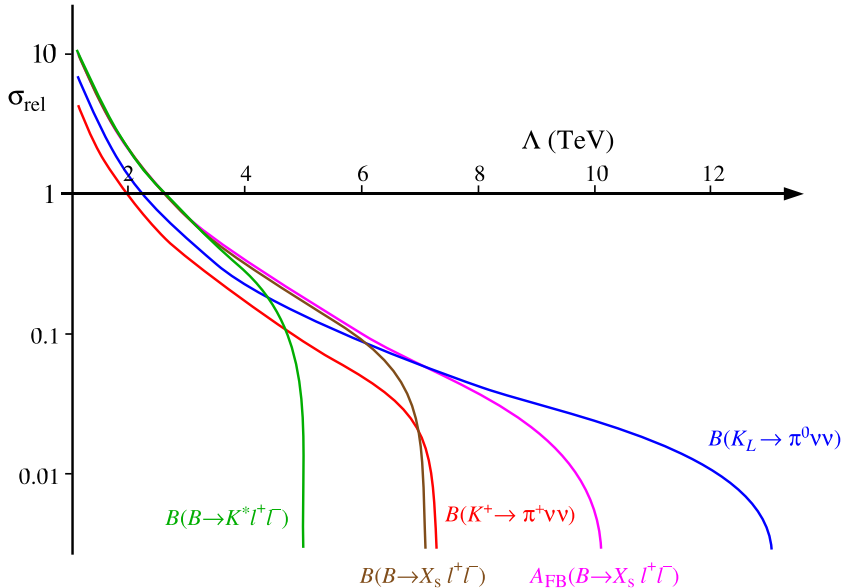}
\end{center}
\vspace{-0.8 true cm}
\caption{\label{fig:MFV} Comparison of the effectiveness 
of different rare observables in setting future bounds 
on the scale of the representative operator 
$(\bar Q_L  Y_U^\dagger Y_U \gamma_{\mu}   Q_L)(\bar L_L \gamma_\mu L_L)$
within MFV models \cite{DGIS}. The vertical axis indicates the relative precision 
of a hypothetical measurement of the observable with central value equal to
the SM expectation.  All the curves are obtained assuming 
a 1\% precision on the corresponding overall CKM factor. }
\end{figure}

An interesting 
virtue of MFV models is that they allow a simple
comparison of new-physics effects in different observables 
in $B$ and $K$ decays. This is because the new contributions
are essentially flavor-universal, with a relative weight in  $B$ and $K$
decays controlled only by the CKM matrix. An  example of this 
comparison is shown in Fig.~\ref{fig:MFV}.
As can be noted, the exceptional theoretical cleanness 
of $\Br(\klpn)$ makes it the most effective probe of 
new physics among rare decays.
 
Other general consequences of the MFV hypothesis, which could 
easily be verified or falsified by precise measurements of 
$\Br(\klpn)$ (or the two $\Kpnn$ rates), are listed below:
\begin{itemize}
\item
The golden relation (\ref{G1}) must be satisfied. As a result,  
given the values of $\sin 2\beta$ and $\Br(\kpn)$, only two values 
of $\Br(\klpn)$ are possible in the full class of MFV models, 
independently of any new detail of the specific framework~\cite{BF01}. 
They correspond to $X$ being positive or negative.  The latter sign
is very unlikely~\cite{MFV05}.
\item
The $95\%$ probability upper bound reads
$\Br(\klpn)\le 4.6\cdot 10^{-11}$~\cite{MFV05}.
\end{itemize}
Apart from these general properties which hold in all MFV models, 
some framework-dependent results, which have been discussed in the 
recent literature, could also be very useful to support or exclude 
specific scenarios: 
\begin{itemize}
\item
within the flavor-blind MSSM~\cite{MSSM}, 
$\Br(\klpn)$ is generally smaller than in the SM; 
\item
within the 
model with one universal extra dimension discussed in~\cite{Edim}, 
one finds $\Br(\klpn)\le 4\cdot 10^{-11}$;
\item
within the so-called littlest-Higgs model, $\Br(\klpn)$
could saturate the $6\cdot 10^{-11}$ bound according to ~\cite{Little}.  On
the other hand, in \cite{DEC05} only deviations from the SM by at most $10\%$
have been found. This discrepancy should be soon clarified.
\end{itemize} 

\subsection{Beyond MFV}
The possibility of new sources of CP violation and flavor mixing 
in the $1-10$ TeV region is, in principle, the most natural
possibility. At present, this scenario is challenged by the precise 
SM-compatible results in $B$ physics. However, a large portion
of the allowed parameter space is still to be explored:
on the one side, it is clear that we cannot have $O(1)$ 
flavor mixing beyond the SM (if new degrees of freedoms 
will show up in the TeV region, as suggested by a natural 
solution to the hierarchy problem); on the other side, 
it is far from being obvious that the SM Yukawa couplings 
are the only source of flavor-symmetry breaking 
(as assumed within the MFV hypothesis). Precise measurements 
of the $\Kpnn$ rates are a key element to address this 
problem in a model-independent and quantitative way.

Models with new sources of CP violation and flavor-symmetry breaking
usually involve large numbers of new free parameters 
that are impossible to fix using only one type of 
experiment. In this case, the information from $\Br(\Kpnn)$
is fully complementary to the information extracted 
by direct searches at high energies, which are crucial 
to determine masses and dominant couplings of the new
particles. The high-energy information is 
not sufficient to fix the (presumably tiny) new effects of CP violation
and flavor mixing: as in the SM, these effects can 
be fully determined only with the help of rare decays.

\begin{figure}[t]
\begin{center}
\hspace{-0.5 true cm}
\includegraphics[width=7.5 cm]{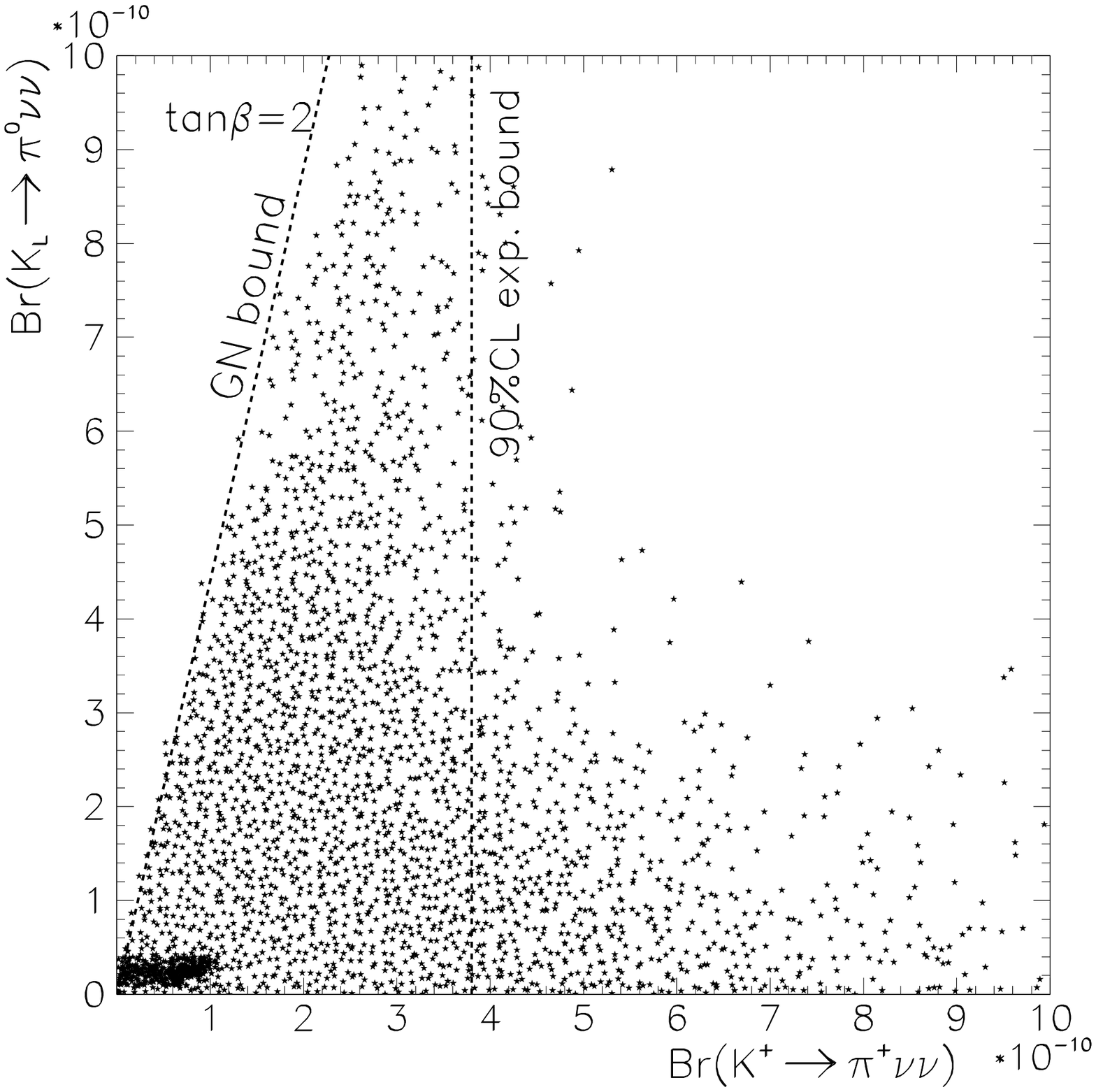}
\hspace{0.1 true cm}
\includegraphics[width=7.5 cm]{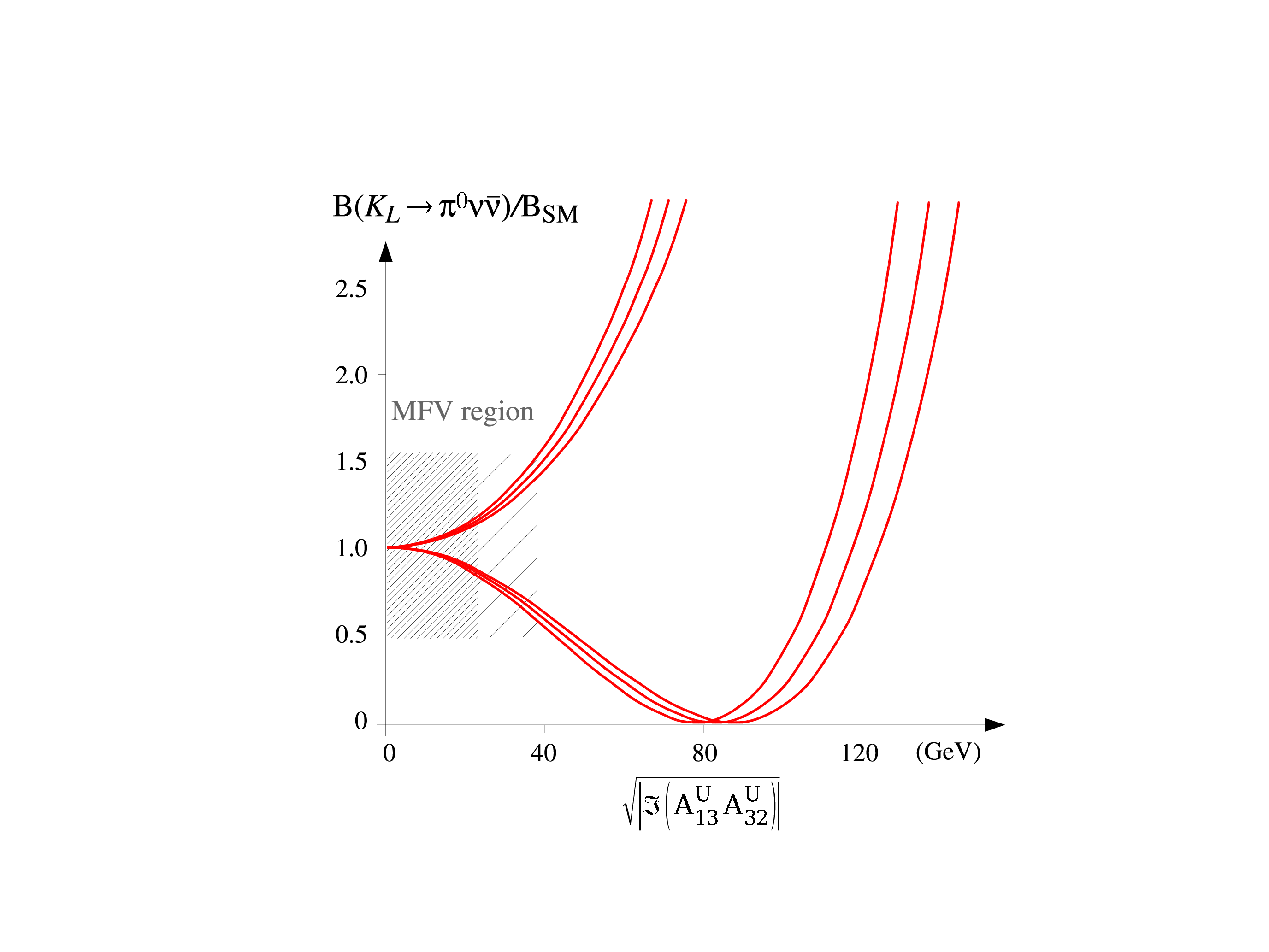}
\end{center}
\caption{Left:  $\Br(\klpn)$ vs.  $\Br(\kpn)$
 as obtained 
by a scan of the allowed parameter space of the MSSM
with generic flavor couplings \cite{GMSSM}. Right: Prediction~\cite{IMST}
of $\Br(\klpn)$ as a function of the soft-breaking trilinear 
couplings $A^U_{13}$ and $A^U_{32}$, at fixed values of squark and 
chargino masses (${\tilde m}_L = 500$~GeV, ${\tilde m}_R = 300$~GeV, 
${\tilde m}_{\chi^\pm} = 200$~GeV, with $\pm 5\%$ uncertainty).
Here  the two branches correspond to the two possible 
signs of the overall MSSM coupling. 
\label{fig:GMSSM}}
\end{figure}

Among the various models of this type which have been 
discussed in the literature, the most representative and 
most popular is probably the MSSM with  generic flavor 
couplings (for a comprehensive analysis of $\Kpnn$ decays 
in this framework, see \cite{GMSSM} and references therein). 
A few important properties which emerge in this context, 
which are also valid in non-supersymmetric models, 
are listed below:
\begin{itemize}
\item{} Even after taking into account all the available 
constraints from CP-violating observables and 
rare decays, there is still much room for possible 
enhancements in  $\Br(\klpn)$ (and also in $\kpn$). The typical range in the MSSM is 
illustrated by the left plot in Fig.~\ref{fig:GMSSM}.
\item{} Large effects in $\Kpnn$ are possible because the 
electroweak structure of the corresponding decay amplitudes 
is quite different from that of $\Delta F=2$ processes
($K^0-\bar K^0$ and $B^0-\bar B^0$) and  
$\Delta F=1$ magnetic transitions
($b\to s\gamma $). As a result, within 
the MSSM $\Kpnn$ amplitudes are strongly 
sensitive to the trilinear soft-breaking terms 
in the up sector, which are poorly constrained by 
other observables~\cite{GI_radcor,BRS,CI}.
As illustrated by the right plot in Fig.~\ref{fig:GMSSM}, 
a precise measurement of $\Br(\klpn)$ would 
provide a very stringent constraint on these 
fundamental couplings of the MSSM, which are
weakly constrained by other sources. 
\item{} The possible values of $\Br(\klpn)$ in this general framework
are not necessarily above the SM prediction: in this context 
it is also possible to obtain a vanishing small $\klpn$ rate
(contrary to the MFV case, where the experimental evidence 
of the $\kpn$ mode also implies a non-vanishing $\klpn$ rate).
\item{} In the presence of new sources of flavor mixing  
 the golden relation (\ref{G1}) is 
naturally broken. 
\end{itemize}
To conclude this section, we note that possible large deviations from the SM 
in the two $\Br(K\to\pi\nu\bar\nu)$ have also been 
discussed recently  in more exotic scenarios, such as supersymmetric models
with broken $R$ parity \cite{DeAndrea}, 
models with extra $Z^\prime$ bosons \cite{HV0}, 
or models with extra vector-like or isosinglet quarks \cite{DESH}. 
A complete list of references can be found in \cite{BSU}.

\boldmath
\section {Experiments Seeking  $\klpn$}
\label{sec:over}
\unboldmath

The  experimental signature for the $\klpn$ decay mode
consists of exactly two photons with the invariant mass of a $\pi^0$,
and nothing else.  The experimental challenge arises from the 34\%
probability that a $K^0_L$ will emit at least one $\pi^0$ in
comparison with the expected decay probability for $\klpn$ which is
ten orders of magnitude smaller. The most difficult decay channel to suppress  is
the CP-violating channel 
$K^0_L\rightarrow \pi^0 \pi^0$, which has a branching ratio of
$0.9\cdot 10^{-3}$ \cite{PDG}. Compounding the problem,
interactions between neutrons and kaons in the neutral beam 
with residual gas in the decay volume can also result in emission of
single $\pi^0$s, as can the decays of hyperons which might occur in the 
decay region, {\it e.g.} $\Lambda\to \pi^0$n. Virtually any experimental
approach must rely on an extremely high level of 
photon detection efficiency, at least as good as the best yet achieved
in E949, the study of $\kpn$, at BNL \cite{E949}. However, due to limitations 
in the level of achievable efficiency due to physical processes such as
photonuclear interactions and pile-up effects,
a firm observation of $\klpn$  at the expected level requires 
some additional handles for suppressing backgrounds.

\begin{figure}[t]
\centerline{
\epsfysize=3.0in
\epsffile{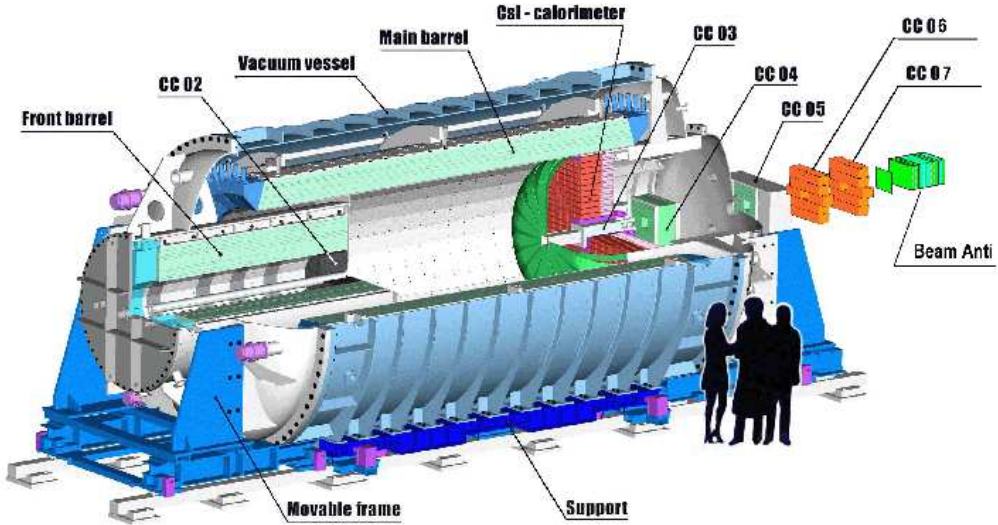}}
\vspace{0.5 cm}
\caption{E391a $K_L \to \pi^0 \nu\bar\nu$ detector at KEK\cite{E391}.  The
neutral beam enters from the left.
}\label{fig:e391a}
\end{figure}

The current experimental limit $\Br(\klpn)~\le 5.9 \times
10^{-7}$ \cite{E799} comes from the KTeV experiment at Fermilab, 
which employed the Dalitz decay $\pi^0\to \gamma e^+ e^-$ with 
charged particles in the final state to obtain a 
better signature  for suppressing backgrounds. 
The two order of magnitude penalty incurred by the $\pi^0\to \gamma e^+ e^-$ 
branching ratio rules out this method for high sensitivity searches.
Considerable
improvement in sensitivity 
is anticipated
by  E391a \cite{E391} 
which has recently taken data at the 12 GeV Proton 
Synchrotron at KEK.
The detector is shown in 
Fig.~\ref{fig:e391a}.
The experiment 
employs a highly collimated 
``pencil'' beam to provide transverse constraints on the
origin of the~$\pi^0$.  The beam enters a cylindrical
veto barrel designed to eliminate background from upstream decays.
Photons from signal $\pi^0$'s decaying in the main barrel
are detected in an array of high-resolution pure CsI modules.  
Lead-scintillator 
shower counters occlude all angles not covered by the CsI 
or the incoming beam, so that there is nearly hermetic veto coverage.
Events with two clusters in the CsI unaccompanied by other
detector activity are fit assuming
they emanate from a $\pi^0$ decaying in the beam.  This allows the
determination of Z-vertex and transverse momentum values for the $\pi^0$.
Cuts on these quantities are designed to distinguish signal from background.
E391a is intended to serve as a pilot for a possible more sensitive
experiment to be mounted at J-PARC\cite{JPAR}.

\begin{figure}[t]
\centerline{
\epsfysize=3.0in
\epsffile{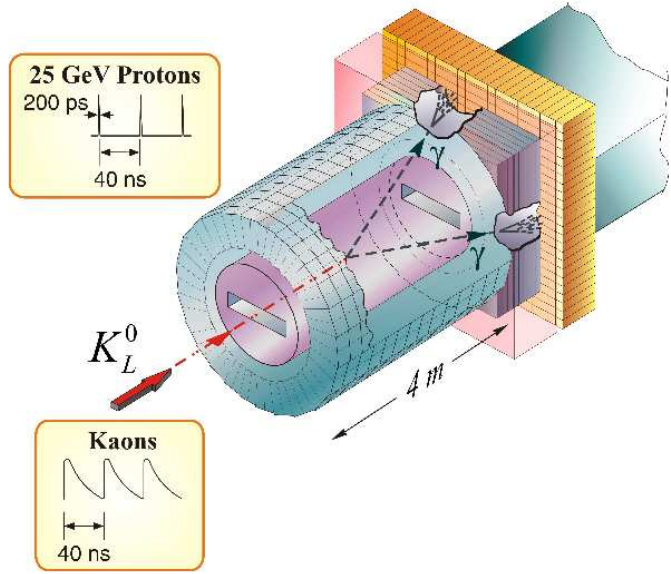}}
\vspace{0.5 cm}
\caption{Schematic  representation of the KOPIO apparatus and technique.
The neutral  kaon beam is
produced by a 25 MHz micro-bunched proton beam striking a production
target. Kaons decaying  via $\klpn$ in the
detector region are detected by the presence of photons from $\pi^0$
decay which convert and are tracked
in the 
photon pointing calorimeter or ``preradiator'' so that the 
$K^0_L$ decay vertex can be determined.
}\label{fig:KOPIO}
\end{figure}

KOPIO is a new experiment at the BNL AGS which seeks to observe and study 
$\klpn$
if it occurs at the SM level or even well below the SM prediction.
The extra handle that  makes a robust experiment feasible is the measurement of 
the  $K^0_L$ momentum using
time-of-flight (TOF) (see Fig.~\ref{fig:KOPIO}). Copious low energy kaons can be produced at the
BNL AGS in an appropriately time-structured beam.  From the knowledge of the
decaying $K^0_L$ momentum, the $\pi^0$ can be transformed to the
$K^0_L$ center-of-mass frame and kinematic constraints can be imposed
on an event-by-event basis
when the $\pi^0$ decay photon directions are measured.
This technique facilitates rejection of
 other kaon decay modes and suppression of all other potential backgrounds,
including otherwise extremely problematic ones such as hyperon decays
and beam neutron and photon interactions.

The required level of background suppression will be 
achieved using a combination of hermetic
high sensitivity photon vetoing and full reconstruction of 
photons through measurements of timing, position, angle, and energy.
Events originating in the two-body decay $K^0_L \to \pi^0 \pi^0$
identify themselves when reconstructed in the $K^0_L$ center-of-mass
system once two photons have been observed.
 Furthermore, those events with missing low energy photons, the
most difficult to detect 
(due, in part, to possible photo-nuclear interactions), 
can be kinematically identified and eliminated.  With the
two  criteria based on precise kinematic measurements and
demonstrated photon veto levels, there is sufficient experimental 
information so that $K^0_L \to \pi^0 \pi^0$ can be suppressed, 
and the background level can also be measured directly from
data.

Evaluation of the KOPIO system leads to the expectation that
$\Br(\klpn)$  could   be measured with a precision of 10\% or better
if the SM prediction holds; this would result in a measurement of 
$|{\rm Im}\lambda_t|<5\%$. If non-SM physics 
results in  a larger rate, as discussed above, the precision on the 
branching ratio would be correspondingly better.

\begin{figure}[t]
\begin{center}
\includegraphics[width=10cm,angle=90]{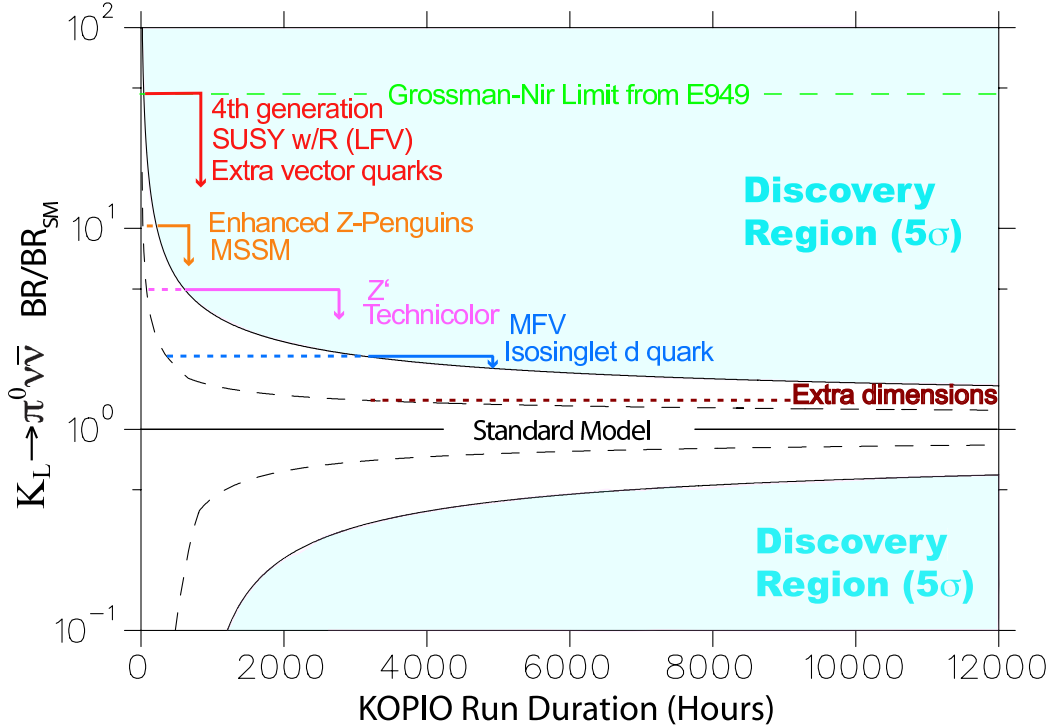}
\end{center}
\caption{5$\sigma$ discovery region (shaded area)
and 95\% CL upper and lower exclusion limits (dashed lines) versus
running time for the KOPIO experiment. For comparison, the maximal 
enhancements of $\Br(\klpn)$ expected in various non-SM 
scenarios (see Section~\ref{sect:BSM})
are also indicated.\label{fraction}}
\end{figure}

As the experimental sensitivity increases, in the absence of a positive signal,
non-SM branching ratios closer and closer to the SM can be eliminated.  
To illustrate the general situation, it is
instructive to use the ultimate reach of an experiment like KOPIO 
where
a  five standard deviation (5$\sigma$) discovery could be firmly
established for branching ratios outside the region
$(0.59 - 1.65) \times \Br(\klpn)_{\rm SM}$.
For shorter runs, the range that can be explored is somewhat smaller
\footnote{For example, after the 6000 hours of operation in the
present plan for KOPIO, a 5$\sigma$ discovery could be made
outside the region $(0.48 - 1.91) \times \Br(\klpn)_{\rm SM}$}.  
A  plot of 
the $5\sigma$ discovery region and of the 95\% CL exclusion
limits as a function of running time is given in Fig.~\ref{fraction}.

Although specific to KOPIO, this figure illustrates an interesting
generic feature of any experiment designed to span orders of magnitude
in searching for a rare process.  First, because the background
rejection power of the experiment must be sufficient for the ultimate
sensitivity, for the early part of the running, the background will be
negligible and progress in ruling out (or discovering!)  branching
ratios far above the expected level is very swift and will be
nearly linear in running time.  After the initial period, to a good
approximation, further progress becomes proportional to the square root
of the running time.  
It is also notable that in any
experiment with a significant amount of background present
along with the signal, to bound the branching ratio from below 
requires a substantial amount of running.
In the KOPIO case, the existence of a SM signal at the
five sigma level would be established after about 1000 hours into the
run.

\section{Conclusions}
The rare decays $\kpn$ and $\klpn$ are both
extremely suppressed within the SM and exceptionally clean
--from the theoretical point of view-- both in the SM 
and in most of its extensions.
For these combined reasons these 
processes play key roles in the search for physics beyond the SM. 
In particular, their measurements offer unique tools to 
deeply investigate the CP violation  and flavor breaking  structure of 
any extension of the SM.  Being completely 
dominated by (one-loop) electroweak dynamics, the two $\Kpnn$ rates
may be greatly affected by  
new-physics contributions. However, even if the experimental 
measurements were found to be in   
agreement with the SM expectation, with a small relative error, 
this information would translate into a precious insight 
about new physics: information about the flavor structure 
of the model complementary to those attainable at high-energy colliders.

Although most of the theoretical virtues are shared by the neutral and 
charged $\Kpnn$  modes, the $\klpn$ channel has the great 
advantage of being sensitive to CP violation and, as a
consequence, of being more sensitive to new physics. In addition, it is 
the  theoretically cleanest of all the accessible FCNC process 
involving quarks. This makes the $\klpn$ mode probably the most 
fascinating process in the field of $K$ and $B$  meson decays. 

Experimentally, the prospects for achieving high precision measurements 
of $\Kpnn$ decays are very promising. E787/E49 at BNL has discovered 
the decay $\kpn$ observing three events so far. 
Initiatives to pursue this measurement are under discussion at 
Fermilab, J-PARC and CERN.   The latter, ``NA48/3'', aims at a sensitivity 
equivalent to a 10\% measurement at  the SM level~\cite{NA48}. Shortly, 
a new result from a recently completed search for $\klpn$ at KEK 
will be available and 
an LOI exists for J-PARC that aims at a high precision measurement\cite{JPAR}. 
The new KOPIO experiment at BNL plans to explore branching ratios well 
below the SM prediction and, in the absence of new physics, would 
measure $B(\klpn)$ to a precision approaching 10\% . This result would 
exclude many possible non-SM approaches and, on pure dimensional grounds, 
would place a limit above 1000 TeV on the mass scale 
of contributing new physics.

\section*{Acknowledgments}

This work was supported in part by Bundesministerium f\"ur
Bildung und Forschung under the contract 05HT4WOA/3 and by the
German-Israeli Foundation under the contract G-698-22.7/2002, 
by the US Department of Energy under contract DE-AC02-98CH1-886,
and by the Natural Sciences and Engineering Research Council and
the National Research Council of Canada.
G.I. wishes to thank the hospitality of the Theory Group at Fermilab,
where part of this work was done, and the partial support by the
IHP-RTN program, EC contract No.~HPRN-CT-2002-00311 (EURIDICE).

\renewcommand{\baselinestretch}{0.95}

\vfill\eject

\end{document}